\documentclass{PoS}

\title{Heavy-light current-current correlators}

\ShortTitle{Heavy-light current-current correlators}

\author{\speaker{Jonna Koponen}, Christine T. H. Davies, for HPQCD\\
        University of Glasgow, UK\\
        E-mail: \email{j.koponen@physics.gla.ac.uk}}

\author{Kent Hornbostel\\
        Southern Methodist University, Dallas, Texas, USA}

\author{Eduardo Follana\\
        Universidad de Zaragoza, Spain}

\author{G. Peter Lepage\\
        Cornell University, Ithaca, New York, USA}

\author{Craig McNeile\\
        Bergische Universit\"at Wuppertal, Germany}

\author{Junko Shigemitsu\\
        The Ohio State University, Columbus, Ohio, USA}

\author{Matthias Steinhauser\\
        Karlsruhe Institute of Technology, Germany}

\abstract{
The current-current correlator method has been used successfully to
obtain very accurate results for quark masses and the coupling $\alpha_s$.
The calculations were done using Highly Improved Staggered Quarks (HISQ) and
heavy-heavy meson correlators. We now extend this work to the significantly
more challenging heavy-light case, reporting the first results here.
The aim is to determine nonperturbative $Z$ factors for NRQCD heavy-light
currents, but first we test the method in the HISQ case where $Z=1$.

}

\FullConference{The XXVIII International Symposium on Lattice Filed Theory\\
                 June 14-19,2010\\
                 Villasimius, Sardinia Italy}

\usepackage{amsmath}

\begin{document}

\section{Motivation}

In the study of semileptonic and leptonic processes, like $B\to \pi l\nu$
and $B\to l\nu$, the nonperturbative $Z$ factors for heavy-light currents
are needed. One way to try to calculate these (nonperturbatively on the
lattice) is the current-current correlator method, that has been successfully
tested and used in the heavyonium case \cite{cbmasses,cmassalpha}.
We now want to extend these results and use the same method in the heavy-light
case. Here we report on the first results using the HISQ action. The eventual
aim is to extract NRQCD heavy-light $Z$ factors.

\section{Current-current correlator method}

The idea is to match time moments of meson correlators to energy-derivative
moments at $q^2=0$ of polarization functions $\Pi$ calculated in continuum
QCD perturbation theory to high order.

The pseudoscalar current-current correlators are defined as
\begin{equation}
G(t)=a^6\sum_{\vec{x}}(am_q)^2\langle 0|j_5(\vec{x},t)j_5(0,0)|0 \rangle .
\end{equation}
Then the time moments are
\begin{equation}
G_n=\sum_t\Big(\frac{t}{a}\Big)^nG(t)
\end{equation}
(see e.g. \cite{cbmasses,cmassalpha}). To help reducing the errors we divide
each moment by the tree level value, $G_n^{(0)}$, and define reduced moments
$R_n$ as
\begin{equation}
\label{Eq_lattRn}
R^{\text{latt}}_4=\frac{G_4}{G_4^{(0)}},\quad\text{and}\quad
R^{\text{latt}}_n=\Bigg(\frac{G_n}{G_n^{(0)}}\Bigg)^{\frac{1}{n-4}}
\quad\text{for}\quad n\geq 6.
\end{equation}
In the continuum the reduced moments are
\begin{equation}
\label{Eq_contRn}
R^{\text{cont}}_4=\frac{g_4}{g_4^{(0)}},\quad\text{and}\quad
R^{\text{cont}}_n=\frac{m_{\eta_h}}{2m_h(\mu)}\frac{g_n}{g_n^{(0)}}
\quad\text{for}\quad n\geq 6.
\end{equation}
The $g_n$ are perturbative series in $\alpha_s(\mu)$, known for the
heavy-heavy case through $\alpha^3_s(\mu)$ \cite{hhpertexp} and for
heavy-light through $\alpha^2_s(\mu)$ \cite{hlpertexp}. The mass $m_h$
is the heavy quark mass in the $\overline{\text{MS}}$ scheme at the
scale $\mu$. Comparing the lattice and continuum $R_n$ allows us to
extract the mass ratio $m_{\eta_h}/(2m_h(\mu))$, and thus the quark mass.
The calculation above is for the case with no $Z$ factor. If the lattice
current has a $Z$ factor then that can also be extracted.

\section{Heavy-light JJ correlators}

\begin{figure}
\centering
\includegraphics*[angle=-90,width=0.648\textwidth]{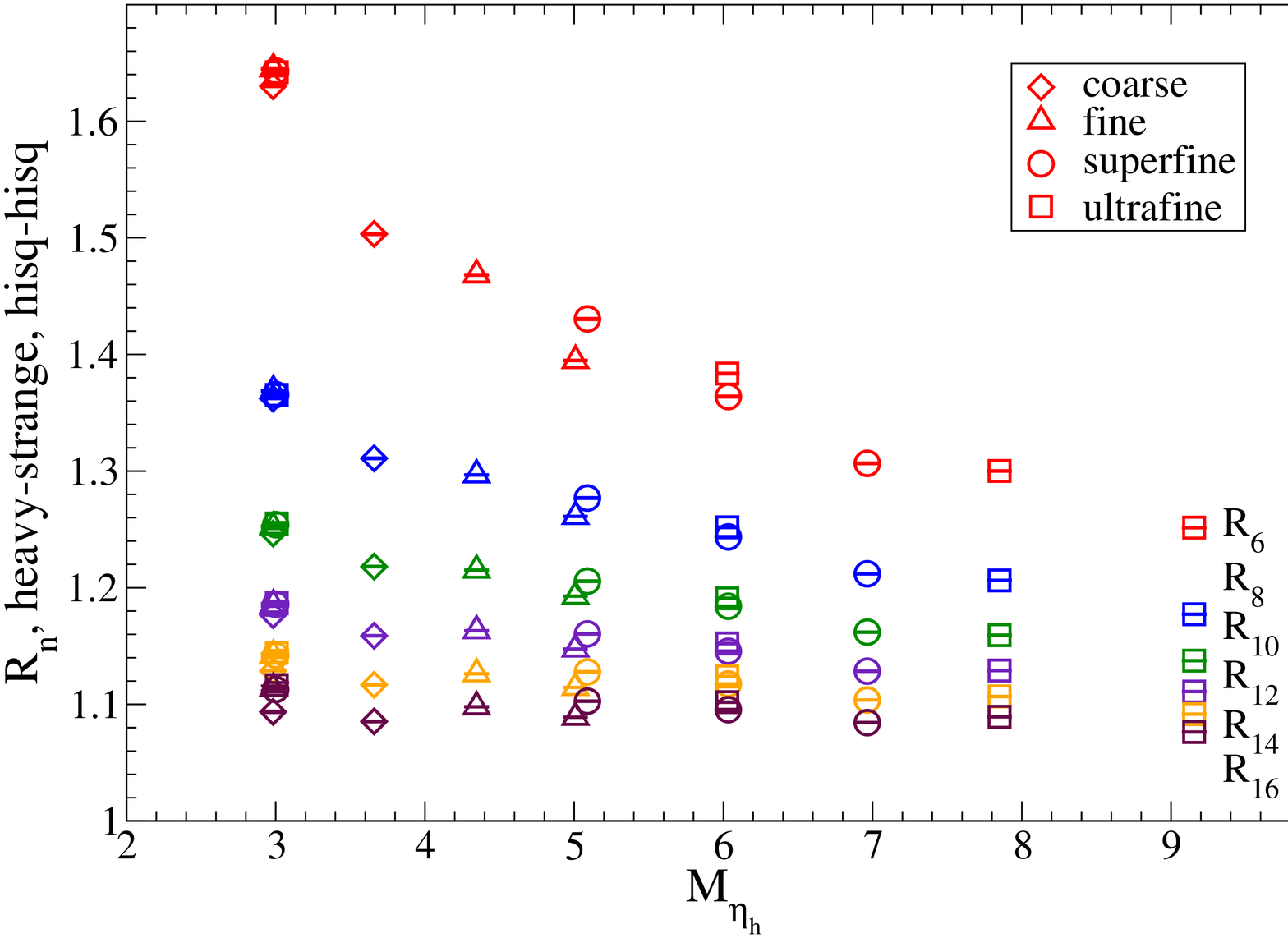}
\includegraphics*[angle=-90,width=0.648\textwidth]{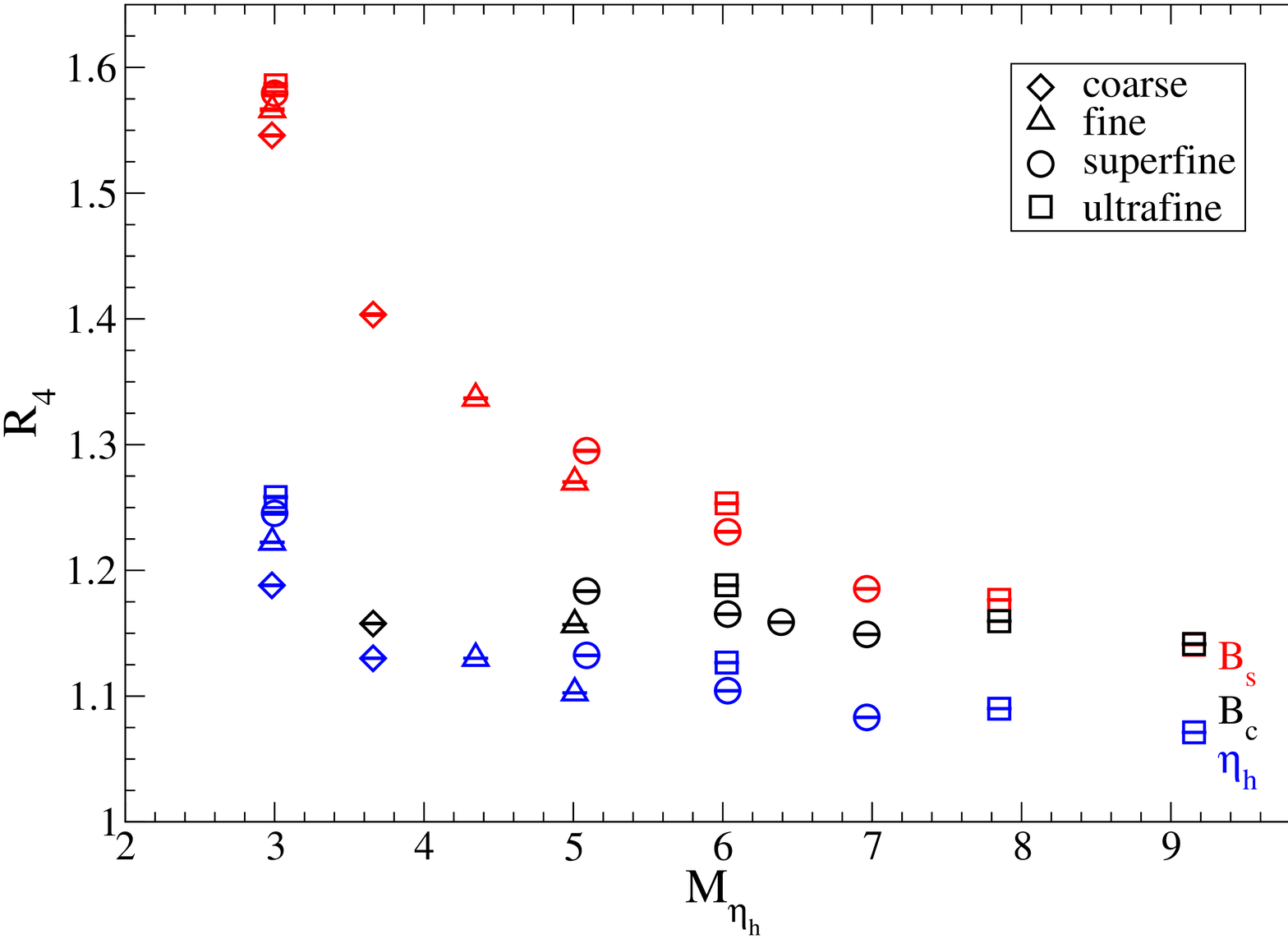}
\includegraphics*[angle=-90,width=0.648\textwidth]{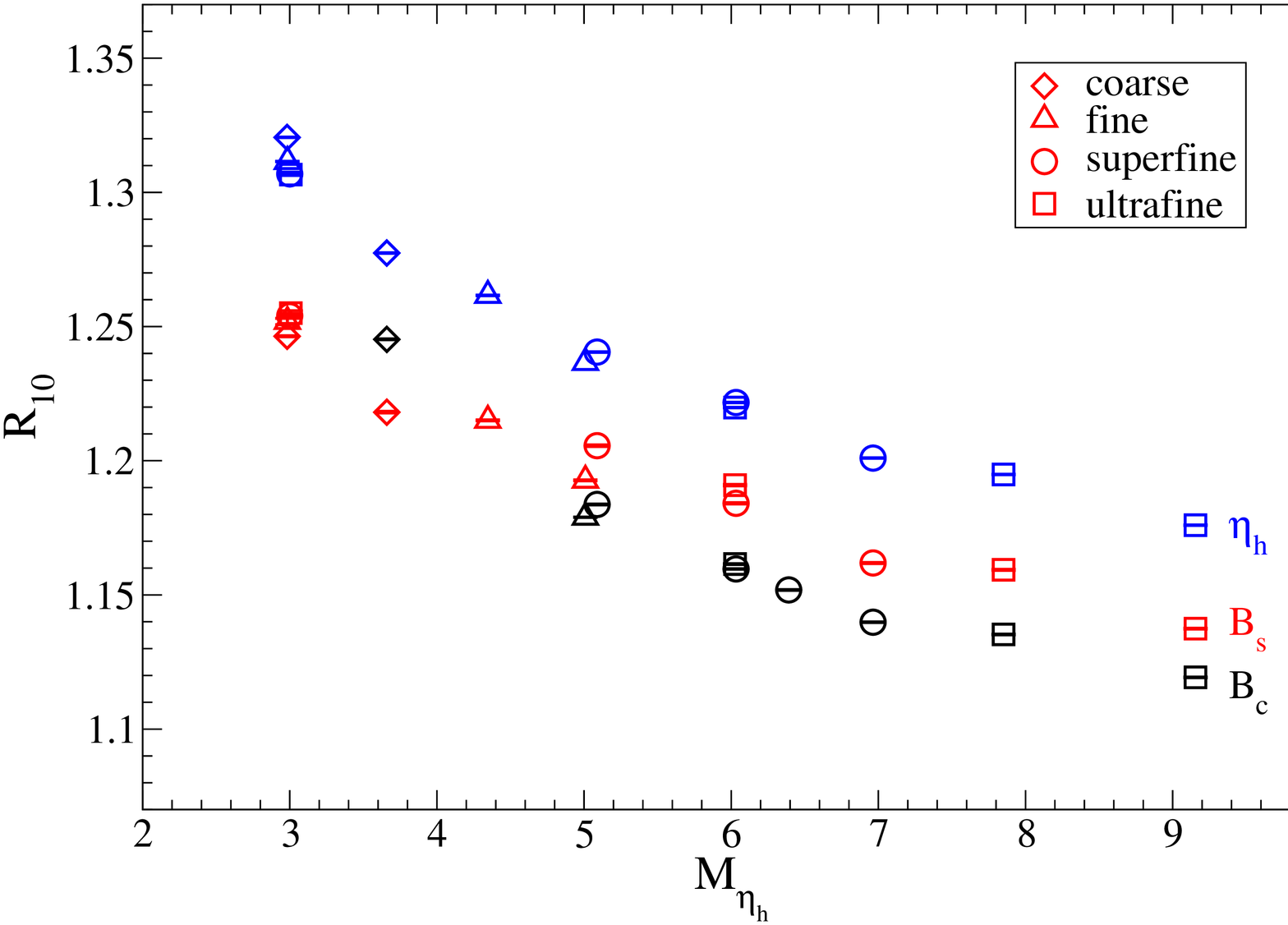}
\caption{Top figure: Heavy-strange correlator reduced moments $R_n$ as a function
of heavy-heavy meson mass $M_{\eta_h}$ (in GeV). The other two figures show the
heavy-strange, heavy-charm and heavy-heavy reduced moments $R_4$ and $R_{10}$ as a
function of $M_{\eta_h}$. The range is from charm (at about 3 GeV) to $b$
(about 10 GeV).}
\label{fig_Rnwrthhmass}
\end{figure}

We compare lattice calculations to continuum perturbation theory through
$\alpha^2_s(\mu)$. In this work we have used coarse, fine, superfine and
ultrafine MILC lattice configurations. We calculate heavy-light correlators
using the HISQ action \cite{hisqpaper} for both quarks with several
heavy quark masses from charm up to the $b$ quark mass. Note that $Z=1$
in the HISQ case. Some of the calculated reduced moments $R_n$ are shown
in Fig.~\ref{fig_Rnwrthhmass} as examples of our results. Comparing
heavy-strange, heavy-charm and heavy-heavy correlator reduced moments
shows that we can clearly distinguish between these three cases.
In the following
subsections we address some challenges of the heavy-light calculations.

\subsection{Volume dependence}

\begin{figure}
\centering
\includegraphics*[angle=-90,width=0.64\textwidth]{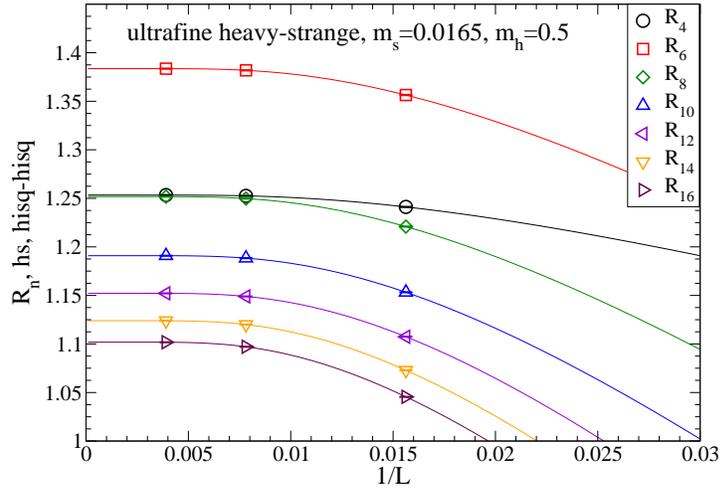}
\caption{The reduced moments in the free, non-interacting theory depend on the
volume.}
\label{fig_volume_free}
\end{figure}

\begin{figure}
\centering
\includegraphics*[angle=-90,width=0.64\textwidth]{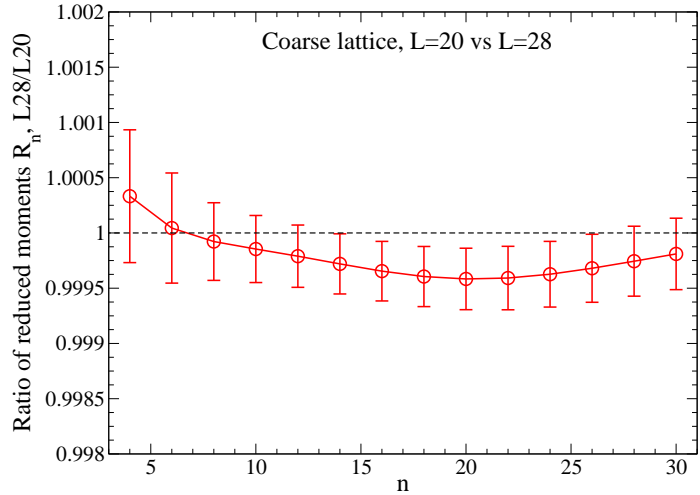}
\caption{The ratio of reduced moments calculated on two different coarse lattices
(one with $L=20$, other one with $L=28$) shows that there is no volume dependence
in the interacting case.}
\label{fig_volume_interacting}
\end{figure}

The tree level (free) moments depend on volume --- note that this is an artifact
of the free case only. This is illustrated in Figures~\ref{fig_volume_free} and
\ref{fig_volume_interacting} --- the $R_n$ depend on volume in the non-interacting
theory, but not in the interacting case. Therefore we need to calculate the tree
level moments in the infinite volume limit. We do this by calculating the free
moments using different volumes, $L^3$, and fitting them with
\begin{equation}
A_0+A_1\frac{e^{-A_2L}}{L}.
\end{equation}
The result in the infinite volume limit is then simply given by the fit
parameter $A_0$.

\subsection{Quark condensate}

\begin{figure}
\centering
\includegraphics*[angle=-90,width=0.64\textwidth]{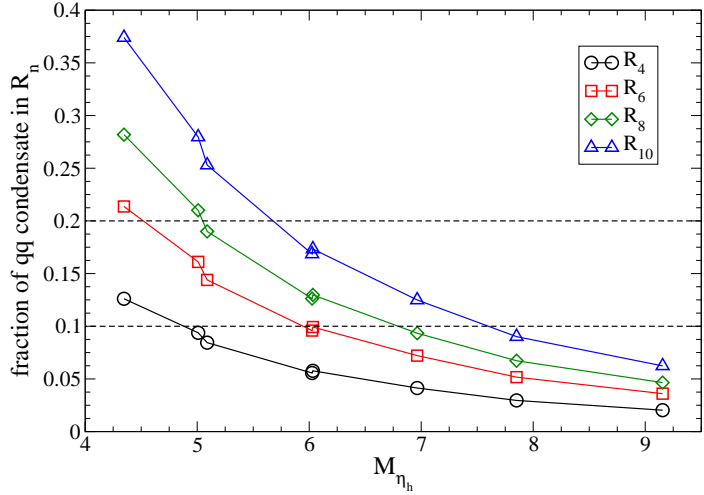}
\caption{Fraction of the tree level $q\bar{q}$ condensate in reduced moment
$R_n$ as a function of the heavy-heavy meson mass $M_{\eta_h}$ (in GeV).}
\label{fig_qqcond}
\end{figure}

The quark condensate appears in the reduced moment $R_n$ at tree level
as \cite{qqcond}
\begin{equation}
\label{eq_qqcond}
\frac{4\pi^2}{3}
\frac{(n-1)(n-2)(n-3)[-\frac{m_h}{m_l}+\frac{n}{2}]}{1+\frac{(n-3)m_l}{m_h}}
\frac{\langle m_l\psi\bar{\psi}\rangle}{m^4_h}.
\end{equation}
The quark condensate is not present in the heavy-heavy case, but it is sizeable in
the heavy-light case --- the fraction of tree level $q\bar{q}$ condensate in $R_n$
can easily be 10--30\% for heavy quark masses masses between $c$ and $b$, as
can be seen in Fig.~\ref{fig_qqcond}. Note that the leading term is $1/m_h^3$.
This poses a challenge, as the $\alpha_s$ corrections to the condensate are
not known. The gluon condensate contribution is much smaller and can be safely
neglected in the analysis.

\subsection{$m_l/m_h$ corrections to perturbative series}

\begin{figure}
\centering
\includegraphics*[angle=-90,width=0.49\textwidth]{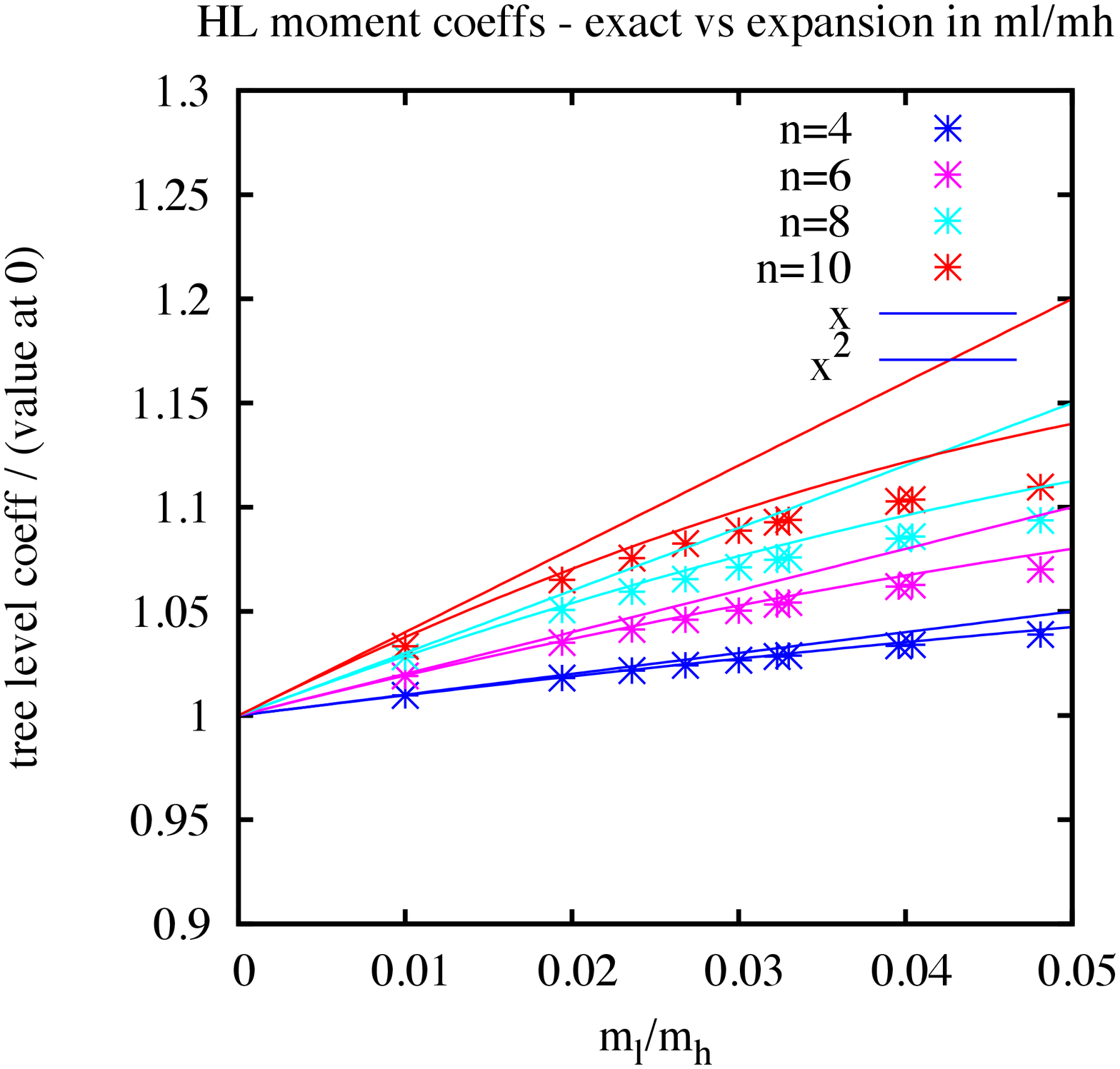}
\includegraphics*[angle=-90,width=0.49\textwidth]{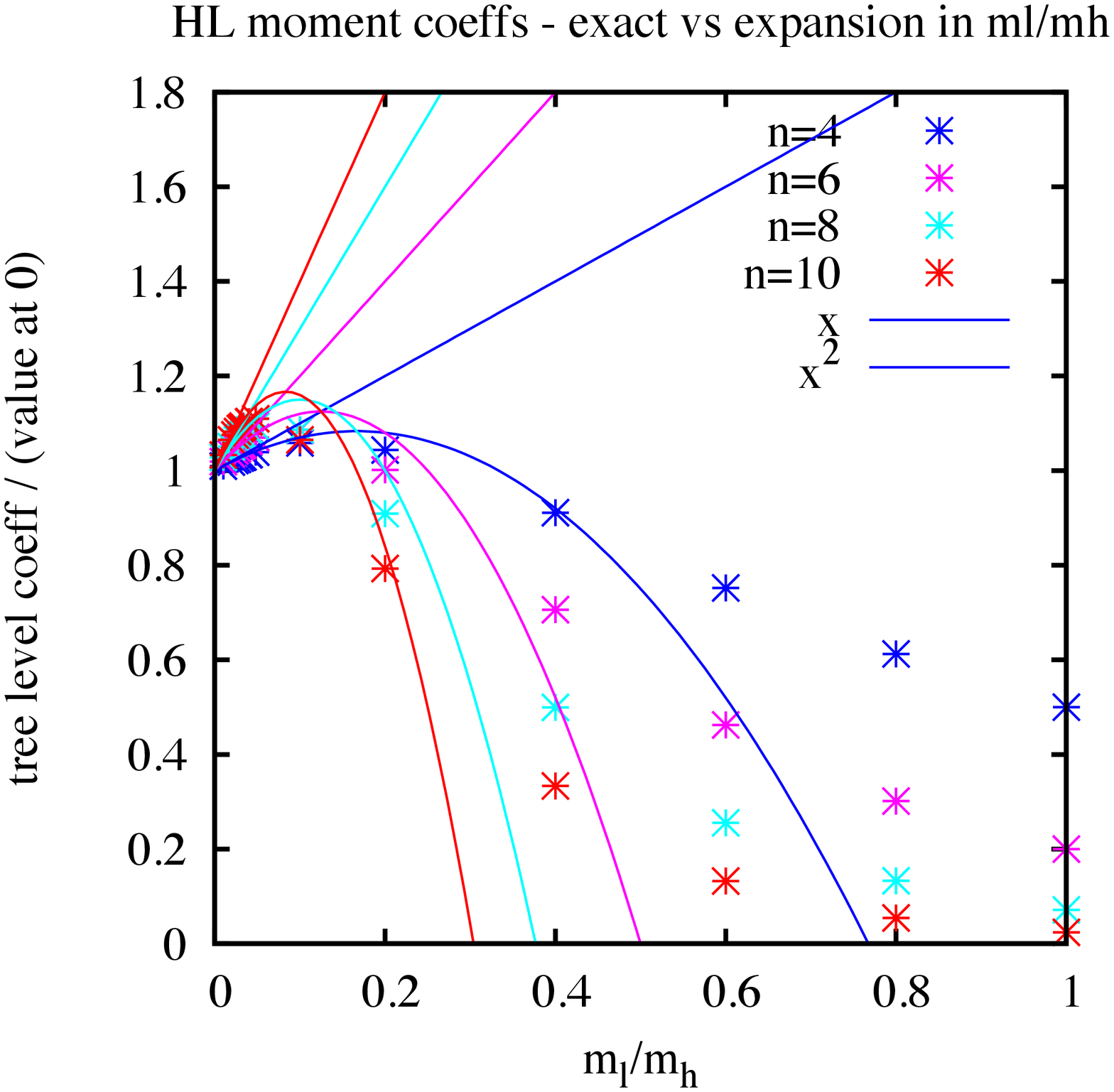}
\caption{$m_l(\mu)/m_h(\mu)$ corrections to continuum perturbation theory
coefficients. Here $x=m_l/m_h$. The tree level values of the coefficients
have been divided by the tree level value at $m_l/m_h=0$.}
\label{fig_massdep}
\end{figure}

Perturbation theory with $m_q=0$ is not sufficient, as $m_l(\mu)/m_h(\mu)$
corrections become important for $B_c$: $m_c(\mu)/m_b(\mu)\approx 0.22$.
At small values of the ratio the $m_l(\mu)/m_h(\mu)$ expansion is good, i.e. it
works for $B_s$. At large values of the ratio the expansion is not good enough.
This is illustrated in Fig.~\ref{fig_massdep}. However, we now have the exact
coefficients for given ratios $m_l(\mu)/m_h(\mu)$ for tree level (shown in
the plot as bursts) and order $\alpha_s$, so this problem can be partly avoided.
The exact coefficients are still needed for $\alpha_s^2$ and higher orders.

\subsection{Fits}

We fit the lattice data $R^{\text{latt}}_n$, $n\geq 6$, with
\begin{equation}
\begin{split}
R_n^{\text{fit}}=&\Big(\frac{m_{\eta_h}}{2m_h(\mu)}\Big)
\big(1+c_1\alpha_s+c_2\alpha_s^2+c_3\alpha_s^3+c_4\alpha_s^4+
c_5\alpha_s^5+c_6\alpha_s^6+q\bar{q}\quad\text{condensate}\big)\\
&\big(1+b_1(am_h(\mu))^2+b_2(am_h(\mu))^4+b_3(am_h(\mu))^6+d_1a^2+d_2a^4\big)
\end{split}
\end{equation}
and choose the scale $\mu=m_h$. We take the first few coefficients ($c_1$ and
$c_2$ in the heavy-light case) from continuum perturbation theory, and treat the
coefficients for the higher order $\alpha_s$ terms as fit parameters. The
quark condensate is given in Eq.~\ref{eq_qqcond} at tree level. We take the
$q\bar{q}$ condensate value to be $\langle m_ss\bar{s}\rangle = (0.2\text{ GeV})^4$
from the Gell-Mann -- Oakes -- Renner relation, allowing the $s\bar{s}$
condensate to be 0.7 times the light quark condensate. We also allow
for the presence of higher order condensate terms estimating them with
powers of the leading condensate. In $B_c$ there is no condensate
contribution, and we get a good fit using the exact coefficients (order
$\alpha_s$). As the lattice calculations are done at a non-zero lattice
spacing $a$, we include $a$-dependent terms in the fit function --- even powers
of $a$ and $am_h(\mu)$. 

To extract the mass ratio $m_{\eta_h}/(2m_h(\mu))$ we use the lattice simulation
data (Eq.~\ref{Eq_lattRn}), with $am_{\eta_h}/am_h$ from the lattice simulations,
and compare these $R^{\text{latt}}_n$ to the continuum perturbation theory result
(Eq.~\ref{Eq_contRn}). That is, we find values for $\alpha_{\overline{\text{MS}}}(\mu)$
and  $m_{\eta_h}/(2m_h(\mu))$ that make lattice and continuum results agree for small
$n > 4$. This can then be combined with experimental results for the $\eta_b$,
$\eta_c$ meson masses to obtain the quark masses. In the heavy-light case we can
use the $\alpha_{\overline{\text{MS}}}(\mu)$ values extracted from the heavy-heavy
calculation.

\begin{figure}
\centering
\includegraphics*[angle=-90,width=0.53\textwidth]{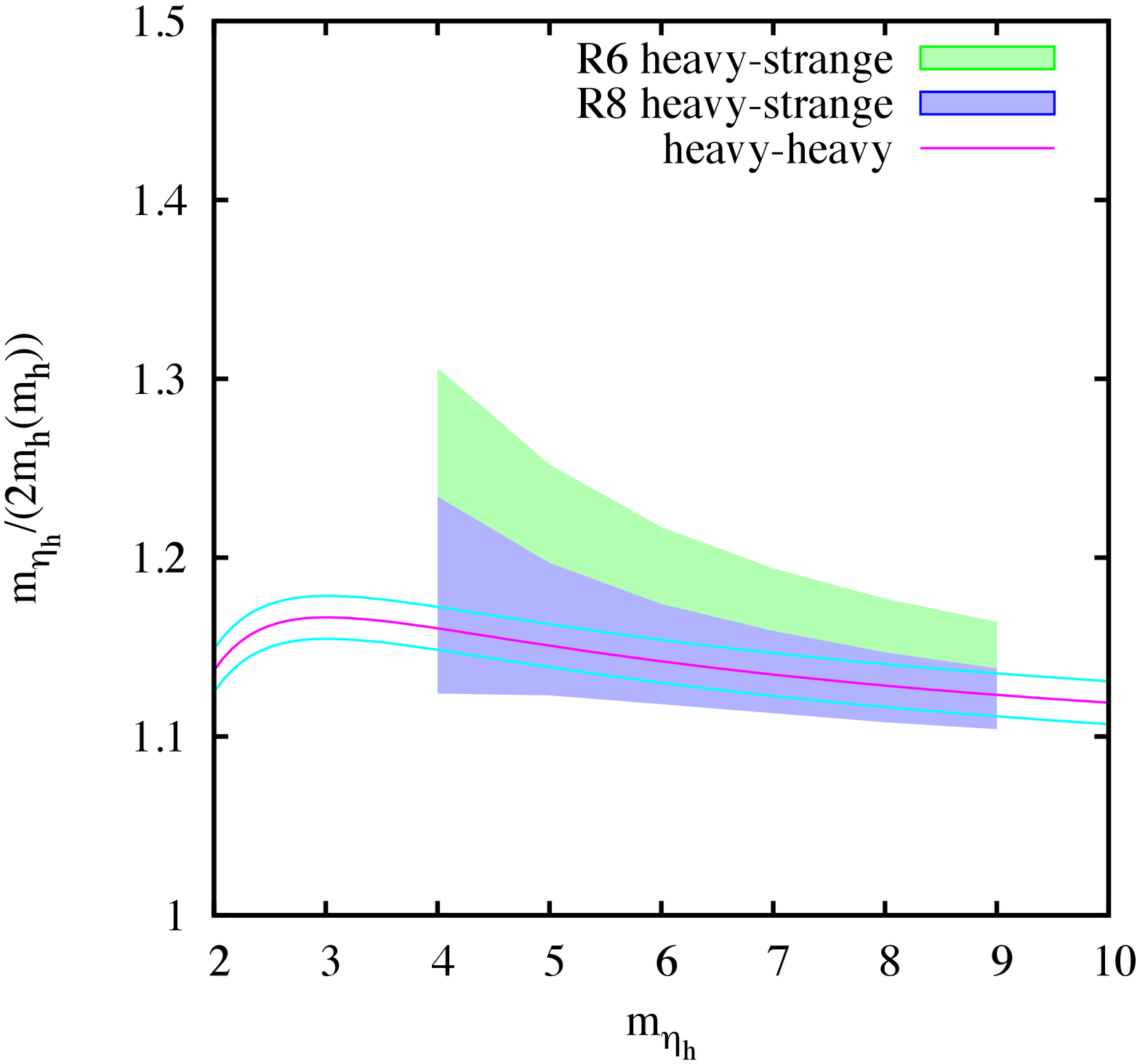}
\caption{The mass ratio $m_{\eta_h}/(2m_h(\mu))$.}
\label{fig_zplot}
\end{figure}

To test the method in the heavy-light case we look at the mass ratio
$m_{\eta_h}/(2m_h(\mu))$ and compare to heavy-heavy results. This is
shown in Figure~\ref{fig_zplot}. In the heavy-strange case the fits are to
one $R_n$ at a time, not to all $R_n$ simultaneously as in the heavy-heavy
case. The mass ratio extracted from the heavy-strange correlator moments
is the same as in the heavy-heavy case, as expected, but currently a lot
less accurate.

\section{Conclusions and future}

We are extending the use of current-current correlator method, earlier used
to study heavy-heavy systems, to heavy-light systems. The full analysis of
heavy-light data is still in progress, but we can already say that the JJ
correlator method works well. The quark condensate sets some limitations ---
we can not use high moments in the $B_s$ case. However, in $B_c$ there is no
condensate contribution. Our aim is to extract $Z$ for NRQCD --- there the
challenge is to control relativistic corrections.

\end{document}